\documentclass[aps,prl,twocolumn,superscriptaddress,showpacs]{revtex4}
\usepackage{graphicx}

\begin{document}


\title{Formation of Hydrogen Impurity States in Silicon and Insulators at Low Implantation Energies}



\author{T.~Prokscha}
\email[E-Mail:]{thomas.prokscha@psi.ch}
\affiliation{Paul Scherrer Institut, Labor f\"ur Myon-Spin Spektroskopie, CH-5232 Villigen PSI, Switzerland}
\author{E.~Morenzoni}
\affiliation{Paul Scherrer Institut, Labor f\"ur Myon-Spin Spektroskopie, CH-5232 Villigen PSI, Switzerland}
\author{D.G.~Eshchenko}
\affiliation{Physik Institut der Universit\"at Z\"urich, CH-8057 Z\"urich, Switzerland}
\affiliation{Paul Scherrer Institut, Labor f\"ur Myon-Spin Spektroskopie, CH-5232 Villigen PSI, Switzerland}
\author{N.~Garifianov}
\affiliation{Kazan Physicotechnical Institute, RAS, Kazan 420029, Russia}
\affiliation{Paul Scherrer Institut, Labor f\"ur Myon-Spin Spektroskopie, CH-5232 Villigen PSI, Switzerland}
\author{H.~Gl\"uckler}
\altaffiliation[Present address:]{Zentralabteilung Technologie, FZ J\"ulich GmbH, D-52425 J\"ulich, Germany}
\affiliation{Paul Scherrer Institut, Labor f\"ur Myon-Spin Spektroskopie, CH-5232 Villigen PSI, Switzerland}
\author{R.~Khasanov}
\affiliation{Paul Scherrer Institut, Labor f\"ur Myon-Spin Spektroskopie, CH-5232 Villigen PSI, Switzerland}
\affiliation{Physik Institut der Universit\"at Z\"urich, CH-8057 Z\"urich, Switzerland}
\author{H.~Luetkens}
\affiliation{Paul Scherrer Institut, Labor f\"ur Myon-Spin Spektroskopie, CH-5232 Villigen PSI, Switzerland}
\author{A.~Suter}
\affiliation{Paul Scherrer Institut, Labor f\"ur Myon-Spin Spektroskopie, CH-5232 Villigen PSI, Switzerland}

\date{\today}

\begin{abstract}
The formation of hydrogen-like muonium (Mu) has been studied as a function of implantation
energy in intrinsic Si, thin films of condensed van der Waals gases (N$_2$, Ne, Ar, Xe), fused
and crystalline quartz and sapphire. By varying the initial energy of positive muons ($\mu^+$) between
1 and 30~keV the number of electron-hole pairs generated in the ionization track of the $\mu^+$ can be
tuned between a few and several thousand. The results show the strong suppression of the formation
of those Mu states that depend on the availability of excess electrons. This indicates, that
the role of H-impurity states in determining electric properties of semiconductors and insulators 
depends on the way how atomic H is introduced into the material.
\end{abstract}


\maketitle

%
%
The implantation of energetic (MeV) positive muons ($\mu^+$) in insulators or
semiconductors commonly leads to the formation of the hydrogen-like bound state
muonium [Mu $= (\mu^+e^-$)] with a final charge state which
can be either positive (Mu$^+$), neutral (Mu$^0$), or negative (Mu$^-$).
In semiconductors Mu is used to identify and investigate the electronic properties
and the behavior of {\em isolated} hydrogen-like states
\cite{patt1988rmp, chow1998, cox2003JPCondMatt},
and hydrogen-related impurities which are of fundamental and technological interest due to
their influence on the electrical and optical properties. Isolated H atoms in materials
are difficult to detect by other spectroscopic means, which mostly
require high H concentrations (see \cite{pereira2005prb, luepke2002prl} and references therein
for examples of vibrational spectroscopy studies of Si).
In contrast, Mu  -- which behaves like a light H isotope ($m_\mu \simeq m_p/9$) --
can be easily detected and characterized by the muon spin rotation ($\mu$SR) technique
due to its high sensitivity per spin.
Muonium states, formed after implantation of energetic $\mu^+$, remain
isolated during the observational time window of the order of the
$\mu^+$ life time (2.2~$\mu$s). Therefore, a large amount of experimental information
on the formation, structure and electrical activity
of isolated H states in semiconductors has been obtained from $\mu$SR, which has
played a pioneering role in the identification and characterization of hydrogen-like
states.
In Si, Ge and semiconductors of the III-V family two Mu states lying deep in the band gap
have been identified at low temperatures ($< 50$~K)\cite{patt1988rmp}: {\em normal} Mu$_{\rm T}$
in the tetrahedral interstitial site with a large isotropic hyperfine interaction (hfi),
and {\em anomalous} Mu$_{\rm BC}$ at a bond-center between two host atoms with a smaller,
anisotropic hfi. In covalent semiconductors, Mu$_{\rm T}$ acts as an acceptor and
Mu$_{\rm BC}$ as a donor.
Recently, novel, very weakly bound Mu states ({\em shallow} Mu, binding energies 15 - 60~meV)
with very low hfi have been established in a number of II-VI and III-V (nitrides) compounds
\cite{gil2001prb, cox2001prl, davies2003apl}. Theoretical work has shown an
universal alignment of hydrogen levels in semiconductors and insulators \cite{vdWalle2003nature},
from which the electronic properties of hydrogen impurities can be derived. The predicted
shallow donor hydrogen states in InN and ZnO have been confirmed experimentally by $\mu$SR
\cite{davies2003apl, cox2001prl}.

However, it has to be kept in mind that by the techniques used so far the spectroscopically
investigated H-isotopes are energetically inserted in the solid. This results in a large
number $N_{eh}$ of electron-hole pairs generated during slowing
down of the incident particle. For instance, all $\mu$SR experiments performed up to now used MeV-$\mu^+$
beams that generate $10^5$ -- $10^6$ electron-hole pairs per implanted $\mu^+$ in the ionization track
\cite{Neh}.
Similar or higher numbers of excess $e^{-}$ are created by the implantation of H or D ions used
in the case of vibrational spectroscopy or channeling experiments. A sizable fraction of these electron-hole
pairs escapes prompt recombination and is still present around the thermalized impurity as shown
in $\mu$SR experiments with applied electric field {\bf E} of both polarities. The
{\bf E}-field clearly changes the Mu formation probability by pushing
track $e^{-}$ and $\mu^+$ apart or together \cite{kra1992prl,sto1997prl,dge1999pla,dge2002prb},
demonstrating that a significant
fraction of Mu in semiconductors and insulators is generated by the capture of
a track $e^{-}$ after the $\mu^+$ has stopped at an interstitial or bond site.
In semiconductors it appears that this so-called {\em delayed} Mu formation (in contrast to
{\em prompt} Mu, where Mu forms during slowing down in charge-exchange cycles,
followed by thermalization of Mu due to elastic collisions \cite{dge2002prb}) is
the origin of Mu$_{\rm BC}$ and for the recently discovered shallow
Mu centers in III-V and II-VI semiconductors \cite{sto1997prl, dge1999pla, dge2003prb}.
The question therefore arises, whether and how the final states are influenced by the
formation process, which is essential for studies 
on technologically important semiconductors and insulators.
This can be studied by using the polarized
low-energy $\mu^+$ (LE-$\mu^+$) beam at the Paul Scherrer Insitute
(PSI, Villigen, Switzerland) \cite{em1994prl, em2003pb} with variable implantation energy
between 1 and 30~keV. It allows to investigate the formation of hydrogen-like Mu
impurity states as a function of energy, {\em i.e.} as a function of  $N_{eh}$.
By varying the energy, $N_{eh}$ can be tuned between a few and several thousand.
This is up to five orders of magnitude less than for conventional MeV-muon beams.
Below 1~keV nearly no track products are generated, thus approximating the case
where H impurities are thermally introduced, which is the way, how trace atoms
are incorporated in the lattice in the course of wafer growth and fabrication
processes.


In this Letter we investigate for the first time
the formation of thermal Mu as a prototype for
isolated H impurities as a function of implantation energy. In addition to
intrinsic Si and sapphire (Al$_2$O$_3$) with more than one type of Mu
 we investigated thin films of van der Waals solids
(s-Ne, s-Ar, s-Xe, s-N$_2$) and fused and crystalline quartz (SiO$_2$)
due to their simplicity concerning the final charge states: only one
type of Mu$^0$ exists with an isotropic hfi close to vacuum Mu$^0$.
We find that {\em delayed} Mu formation is energy dependent
in the keV range in all the investigated samples.
Below $\sim$~10~keV the formation of those H impurity states
that require a sizeable amount of excess $e^-$ is strongly suppressed.
The data on Si and Al$_2$O$_3$ support the interpretations that
Mu$_{\rm BC}^0$ in Si \cite{sto1997prl} and Mu$^-$ in Al$_2$O$_3$
\cite{brewer-jd2000} are formed by {\em delayed} capture of a track $e^-$.

The $\mu$SR technique allows to differentiate between paramagnetic
(Mu$^0$) and $\mu^+$ in diamagnetic environment
(free $\mu^+$, Mu$^+$ or Mu$^-$). Due to the hyperfine coupling
between the $\mu^+$ and the $e^{-}$ spin the observable Larmor precession frequency
of isotropic Mu$^0$ is about 103 times larger than for the free $\mu^+$. It
splits into two intra-triplet lines that merge to one line at low fields
($< 2$~mT) where 50\% of the muon polarization is not observed due to unresolved hyperfine oscillations
between the triplet and singlet state. The diamagnetic and paramagnetic
decay asymmetries $A_{D}$ and $A_{\rm Mu}$ were determined by measuring the
amplitudes of the $\mu^+$ and Mu precession signals in transverse (perpendicular to the $\mu^+$ spin)
magnetic field, applied parallel to the sample normal.
$A_{D}$ and $A_{\rm Mu}$ are proportional to the fraction of muons in that particular states.

\begin{figure}
\includegraphics[width=1.0\linewidth]{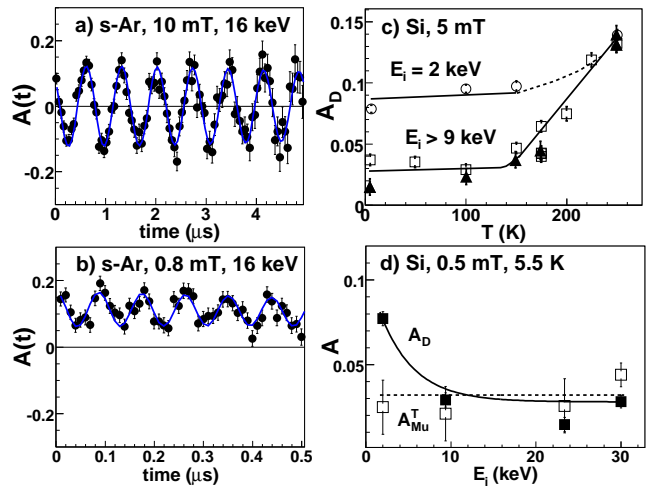}%
 \caption{\label{fig1}
 a) Typical $\mu$SR asymmetry spectrum $A(t)$ for the diamagnetic signal
 in s-Ar, and b) corresponding signal at low fields showing
 the 103-times faster Mu precession superposed to the slow diamagnetic signal. 
 c) Undoped Si, diamagnetic 
 asymmetry $A_{D}$ as a function of temperature $T$. Solid triangles:
 implantation energy $E_{i} > 20$~keV, open squares:
 $E_{i} = 9.3$~keV, open circles: $E_{i} = 2.0$~keV.
 d) $A_D$ and Mu$_{\rm T}$ asymmetry
 $A_{\rm Mu}^{\rm T}$ as a function of $E_{i}$.
 The lines in a) and b) are fits, and in c) and d) guides to the eye.
}
\end{figure}
The 0.5-mm thick Si sample with 50~mm diameter
(undoped, resistivity 10 k$\Omega$cm, capped by a 2-nm thick oxide layer)
was oriented with the $\langle100\rangle$ direction parallel to the sample normal.
The quartz disc samples had thicknesses of
1 and 2~mm [SiO$_2$ crystal and fused quartz ({\tt Suprasil}), respectively]
and 50~mm diameter. The Al$_2$O$_3$ sample was a 0.5 mm-thick single
crystal with 60 mm diameter.
The solid gas films were grown at partial pressures between
$10^{-6}$ and $5\times 10^{-5}$~hPa. Film thicknesses
were about 1000~nm which is sufficient to stop all LE-$\mu^+$ in
the layer. At these deposition pressures grain sizes of order
100~nm are obtained \cite{rgs1975}.
For details on the experimental setup we refer to ref.~\cite{tp2003pb}.

\begin{figure}
 \includegraphics[width=0.75\linewidth]{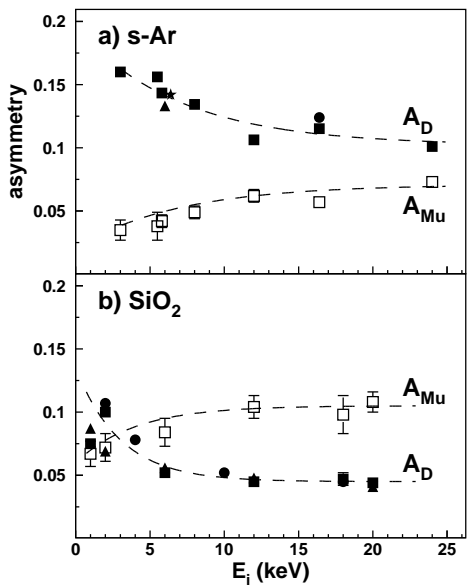}%
 \caption{\label{fig2} Muon and Mu asymmetries
 $A_{D}$ and A$_{\rm Mu}$ as a function of  implantation energy $E_{i}$ for a) s-Ar grown at
 $6.5\times 10^{-6}$~hPa, and b) SiO$_2$ crystal and glass ({\tt Suprasil}), $T = 20$~K.
 The magnetic fields are 10~mT (circles),
 5~mT (triangles), 2~mT (stars) and 0.8~mT (squares). The lines are guides to
 the eye.
}
\end{figure}

Figure~\ref{fig1} shows typical $\mu$SR asymmetry spectra, and displays the results for Si. 
In Si, at 5~mT only the precession of the diamagnetic
signal is observed. Due to limited statistics and time resolution Mu precession
frequencies $> 30$~MHz are too high to be resolved with our present setup.
In a field of 5~mT, the Mu$_{\rm T}^0$ intra-triplet lines are at about 70~MHz, whereas
the Mu$_{\rm BC}^0$ transitions are between 35 and 50~MHz, depending on the orientation of
the {\bf B}-field with respect to the $\langle111\rangle$
crystal axis. At 0.5~mT the Mu$_{\rm BC}^0$ frequencies are nearly unchanged and therefore
not observable with our setup, whereas the 7-MHz signal of Mu$_{\rm T}^0$ becomes visible.
The 0.5-mT data are fitted with two components, a $\mu^+$ precession signal and the
Mu$_{\rm T}^0$ signal with exponential relaxation, whereas the 5-mT data
are fitted with the $\mu^+$ precession signal only.
The temperature dependence of $A_{D}$ at different implantation energies $E_{i}$ is shown in
Fig.~\ref{fig1}a). Above 9~keV - corresponding to a mean implantation depth $\langle d\rangle$
of 65~nm and $N_{eh} \simeq 2400$ \cite{Neh} - $A_{D}$ exhibits the
same temperature behavior as in bulk $\mu$SR experiments \cite{sto1997prl}.
The increase of $A_{D}$ above $\sim$~150~K reflects the thermally induced
ionization of Mu$_{\rm BC}^0$. At $E_{i} = 2$~keV ($\langle d\rangle\sim 18$~nm)
and $T < 150$~K $A_{D}$ is significantly larger than at higher energies.
As Fig.~\ref{fig1}b) shows the behavior of $A_{D}$ is
not related to a change of the Mu$_{\rm T}^0$ fraction, which contrary to $A_{D}$
does not depend on $E_{i}$. It rather reflects the unobserved Mu$_{\rm BC}^0$ fraction
which decreases with decreasing energy and number of available track $e^-$. This is
also supported by the convergence of the two curves in Fig.~\ref{fig1}a) at high $T$
where Mu$_{\rm BC}^0$ is ionized. The $\sim 2$-nm thick oxide layer present on the
Si surface is too thin to explain the observed reduction of the Mu formation.
Moreover, in a SiO$_2$ layer  at low energy a higher Mu fraction should be
observed, see Fig.~\ref{fig2}b). The different dependence
on the availability of excess $e^{-}$ indicate that the main fraction of
Mu$_{\rm BC}^0$ is due to {\em delayed} formation, whereas Mu$_{\rm T}^0$ is
a consequence of charge-exchange processes at epithermal energies --
in agreement with bulk $\mu$SR studies, where an applied {\bf E}-field
was used to vary the average distance between $\mu^+$ and excess $e^{-}$,
and therefore the relative formation probability of these two states \cite{sto1997prl}.

\begin{figure}
 \includegraphics[width=0.8\linewidth]{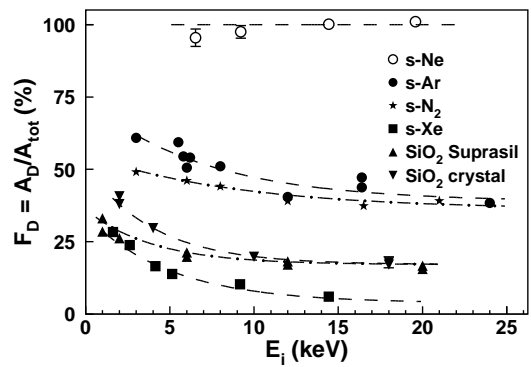}%
 \caption{\label{fig3}
 Comparison of the diamagnetic fraction
 $F_{D}$ as a function of implantation energy $E_{i}$ for different samples,
 $B = 10$~mT.
 Deposition pressures were:
 $7\times 10^{-6}$~hPa for s-Ne,
 $6.5\times 10^{-6}$~hPa for s-Ar,
 $2.2\times 10^{-5}$~hPa for s-N$_2$, and
 $1.5\times 10^{-5}$~hPa for s-Xe.
 The lines are guides to the eye.
}
\end{figure}

Figure~\ref{fig2} shows the energy dependence of $A_{D}$ and $A_{\rm Mu}$ for
s-Ar (a) and SiO$_2$ (b). Only isotropic Mu is present, and
$A_{D}$ and $A_{\rm Mu}$
represent a direct measure of the $\mu^+$ and Mu fraction in the sample.
The sum $A_{tot}$ = $A_{D}+2A_{\rm Mu}$~=~0.263(1) is the total observable asymmetry, and
there is within the experimental errors no missing fraction. Qualitatively,
the s-Ar and SiO$_2$ data display the same behavior: with increasing energy $A_{D}$ is
decreasing while $A_{\rm Mu}$ is increasing correspondingly.
The energy-dependent diamagnetic fractions $F_{D} = A_{D}/A_{tot}$ for various insulators
are summarized in Fig.~\ref{fig3}. With the exception of s-Ne all samples show a
decreasing diamagnetic fraction with increasing energy. For SiO$_2$ and s-Xe bulk Mu
fractions $F_{\rm Mu } = (1-F_{D}$) of 85\% and $\sim 100$\%, respectively,
are obtained at $20$~keV [corresponding to $\langle d\rangle = 155$~nm
(SiO$_2$), $\langle d\rangle = 185$~nm (s-Xe)]. At this energy the number
of electron-hole pairs created in the ionization track is about 1000 \cite{Neh}.
In the s-Ar and s-N$_2$ films even at the highest energy the observed
Mu fractions ($F_{\rm Mu} \sim 60\%$) are lower than the bulk results obtained with
4-MeV $\mu^+$ [$F_{\rm Mu}\sim 100\%$ (s-Ar), $F_{\rm Mu}\sim 80\%$ (s-N$_2$ at $T < 30$~K)].
The discrepancy is even more drastic for s-Ne where the film data are consistent
with $F_{\rm Mu} = 0$ in contrast to the bulk data with $F_{\rm Mu} = 90\%$ \cite{dge2002prb}.
This disagreement can be explained by the suppression of Mu formation in granular s-Ne, s-Ar
and s-N$_2$ thin films, as we discuss below.

The decrease of $F_{D}$ with increasing $E_{i}$ reflects the onset of {\em delayed} Mu formation
with increasing availability of excess $e^-$. From the flattening of $F_{D}$ at $\sim 20$~keV
we estimate the number of excess $e^-$ necessary to saturate the {\em delayed} Mu yield to
be of the order of thousand. The $e^-$ may escape recombination with the $\mu^+$ by several
processes: recombination with a cation from the ion track, trapping at grain boundaries,
voids, and surfaces or escape from the surface ($e^-$ escape depth $\sim 100$~nm in
s-Ar and s-Xe \cite{bar1998prb}, $e^-$ mean free path in Si is $\sim 20$~nm at 300~K, increasing
to $> 100$~nm at lower $T$). An additional obstacle for electron-muon recombination is also the
large escape depth of 20 - 100~nm of {\em epithermal} $\mu^+$ in wide band gap insulators such as
s-N$_2$, s-Ar and s-Ne \cite{em2004JPCondMatt}: after leaving the charge-exchange cycles where
the last $e^{-}$ are released the $\mu^+$ may move such a distance away from
its ionization track, further losing energy inefficiently by elastic collisions.
This large $e^- - \mu^+$ separation and the trapping of $e^{-}$
combine all together to suppress the {\em delayed} Mu formation channel in s-Ne, s-Ar and s-N$_2$.
The total suppression of Mu formation in s-Ne is probably a consequence
of a  $\mu^+$ escape depth larger than the typical grain size, making the formation of
a {\em delayed} $e^- - \mu^+$ bound state unlikely.

\begin{figure}
 \includegraphics[width=0.9\linewidth]{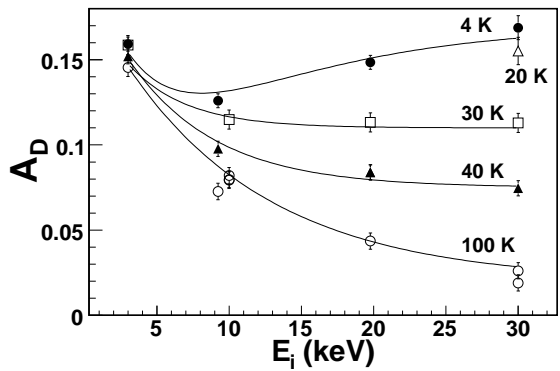}%
 \caption{\label{fig4}
 Diamagnetic asymmetry $A_{D}$
 as function of implantation energy $E_{i}$ for
 sapphire, measured at $B = 10$~mT
 at different temperatures. The lines are guides to the eye.
}
\end{figure}

The energy dependence at $T < 100$~K of $A_{D}$ in sapphire (Fig.~\ref{fig4})
shows an interesting anomaly compared to the data presented so far.
At 100~K $A_{D}$ decreases with increasing energy and reaches
its smallest value of 0.025 at 30~keV. This behavior correlates with the onset of {\em delayed}
formation of Mu$^0$ as seen in other insulators. The energy dependence of $A_{D}$ becomes
less pronounced on reducing the temperature. At 4~K $A_{D}$ exhibits a minimum at 10~keV and
starts to increase again when further increasing the energy. This may reflect the
{\em delayed } formation of diamagnetic Mu$^-$, as suggested in a
previous {\bf E}-field $\mu$SR experiment where the disappearence of Mu$^-$ with
increasing $T$ is interpreted as thermal ioniziation of Mu$^-$ with an activation
temperature of 130~K \cite{brewer-jd2000}.
A recent theoretical work shows that H$^-$ could be the stable charge state
in Al$_2$O$_3$ \cite{peacock2003}. Our data support this idea, and that
Mu$^-$ is formed by {\em delayed} $e^-$ capture.

In conclusion, the measured energy dependence of Mu formation in intrinsic Si
and insulators shows as a general behavior that the formation of {\em delayed}
Mu states requires the presence of the order of thousand excess $e^{-}$ in the
ionization track. With LE-$\mu^+$ H-impurity states can be studied without the 
generation of a non-equilibrium electron track. 
From the implantation energies involved we infer that the length
scale of that part of the track that is involved in {\em delayed} Mu formation is
of the order of 100~nm. At energies $<$~3~keV {\em delayed} Mu formation is nearly absent.
This indicates that the formation of those H-impurity states which heavily depend on the
availability of excess $e^{-}$ is strongly suppressed in cases where the H-isotope is
inserted in the solid without the concomitant presence of a sizeable number of excess $e^{-}$.
This implies, that the role of H-impurity states in determining electric properties of 
semiconductors and insulators depends on the way how atomic H is incorporated into the 
material.
%
The question of the relative importance of different possible 
H-states and their occurrence as native impurity states in 
semiconductors and insulators is generally not addressed and 
we hope that our results will foster new theoretical and experimental 
studies in this area.

We are extending this kind of experiments to the investigation of shallow Mu
states: 
Preliminary data in ZnO show also a decreasing shallow Mu fraction with
decreasing implantation energy, which further supports the general trend
found in other materials.

This work was fully performed at the Swiss Muon Source S$\mu$S, Paul Scherrer
Institute, Villigen, Switzerland. We are grateful to S.F.J. Cox for valuable
discussions. We thank C. David from LMN at PSI for providing the Si sample.
We thank M. Birke, Ch. Niedermayer and M. Pleines for their help in the
initial phase of the experiment. The technical support by H.P. Weber is
gratefully acknowledged.
\bibliographystyle{prsty}

\end{document}